\documentclass[sigconf,authorversion,screen]{acmart}

\usepackage{graphicx} 
\usepackage{comment}
\usepackage{tabularx}
\usepackage{makecell}

\usepackage{listings}

\lstdefinelanguage{JavaScript}{
    keywords={typeof, new, true, false, catch, function, return, null, switch, var, if, in, while, do, else, case, break, cy, Cypress, length, \$body, Commands, const, RegExp, to, let, setTimeout, button, class, Action},
    keywordstyle=\bfseries,
    ndkeywords={export, boolean, throw, implements, import, this},
    ndkeywordstyle=\bfseries,
    morekeywords={[2]{get, click, type, then, find, wait, log, waitUntil, trim, should, expect, eq}},
    keywordstyle=[2]\bfseries,
    identifierstyle=,
    sensitive=false,
    comment=[l]{//},
    morecomment=[s]{/*}{*/},
    commentstyle=\ttfamily,
    stringstyle=\ttfamily,
    morestring=[b]',
    morestring=[b]"
}

\lstdefinelanguage{ShellOutput}{
    keywords={},
    sensitive=false,
    stringstyle=\ttfamily,
    morestring=[b]',
    morestring=[b]",
    morekeywords={[2]{cy}},
    keywordstyle=[2]\bfseries,
    morekeywords={[3]{get, should, click, find}},
    keywordstyle=[3]\bfseries,
    morekeywords={[4]{failed}},
    keywordstyle=[4]\bfseries,
    morekeywords={[5]{passed, successfully}},
    keywordstyle=[5]\bfseries,
    morekeywords={[6]{Repository, Commit, Cloning, Extracting, Setting, Replacing, Running, Locator, File, Line, Getting, Docker}},
    keywordstyle=[6]\bfseries,
    morekeywords={[7]{Dockerfile, run\_tests}},
    keywordstyle=[7]\bfseries,
}

\usepackage{booktabs}
\usepackage{multirow}
\usepackage{xcolor}

\usepackage{listings}

\renewcommand{\texttt}[1]{\text{\small\ttfamily#1}}

\title{ReproBreak: A Dataset of Reproducible Web Locator Breaks}

\author{Thiago Santos de Moura}
\orcid{0009-0007-1927-8378}
\affiliation{%
  \institution{Ruhr University Bochum}
  \city{Bochum}
  \country{Germany}
}
\email{thiago.santosdemoura@acm.org}

\author{Leon Adamietz}
\orcid{0009-0007-6581-6552}
\affiliation{%
  \institution{Ruhr University Bochum}
  \city{Bochum}
  \country{Germany}
}
\email{leon.adamietz@rub.de}

\author{Samra Mehboob}
\orcid{0009-0007-0525-2599}
\affiliation{%
  \institution{Ruhr University Bochum}
  \city{Bochum}
  \country{Germany}
}
\email{samra.mehboob@acm.org}

\author{Yannic Noller}
\orcid{0000-0002-9318-8027}
\affiliation{%
  \institution{Ruhr University Bochum}
  \city{Bochum}
  \country{Germany}
}
\email{yannic.noller@acm.org}

\begin{abstract}
Automated GUI testing frameworks such as Cypress and Playwright rely on locators to find and interact with web elements. A locator break occurs when a structural change in the application under test causes a locator to no longer find its target element, resulting in test breakages even when the underlying functionality remains unchanged. Despite its impact on test maintenance, no dataset exists to evaluate locator fragility in Cypress and Playwright at scale. In this paper, we present ReproBreak, a dataset of reproducible locator breaks in web GUI tests. We analyzed 359 open-source repositories to identify commits that contain locator changes. To confirm whether these changes are indeed locator breaks, we reproduced them in the top 4 projects with the largest number of locator changes and found 449 locator breaks, which are provided in the dataset along with scripts for automated reproduction. We believe ReproBreak serves as a valuable artifact to support research on locator fragility, repair techniques, and test robustness. The video is available at: \url{https://youtu.be/fM6qQhoTjqY}. The dataset is at \url{https://github.com/rub-sq/ReproBreak}.
\end{abstract}



\copyrightyear{2026}
\acmYear{2026}
\setcopyright{cc}
\setcctype{by}
\acmConference[ASE '26]{Proceedings of the 41st IEEE/ACM International Conference on Automated Software Engineering}{October 12--16, 2026}{Munich, Germany}
\acmBooktitle{Proceedings of the 41st IEEE/ACM International Conference on Automated Software Engineering (ASE '26), October 12--16, 2026, Munich, Germany}
\acmDOI{10.1145/3832783.3834638}
\acmISBN{979-8-4007-2882-2/2026/10}

\begin{document}

\maketitle

\section{Introduction}\label{sec:intr}
Modern web applications are present across most software systems, ranging from e-commerce platforms to public services. As these applications evolve rapidly, the risk of functional and visual regressions rises, making reliable Graphical User Interface (GUI) testing increasingly important. Frameworks such as Cypress \cite{cypress2026} and Playwright \cite{playwright2026} are widely adopted in the industry to automatically simulate user interactions, verify GUI behaviors, and detect regressions \cite{di2025investigating}. Unlike unit tests, GUI tests validate the system as a whole and detect integration gaps in a real environment, which is especially relevant for complex web applications.
%
Although these frameworks are effective, the tests written and executed with them rely on element locators, e.g., IDs, CSS selectors, or XPath expressions, to identify and interact with GUI elements. When the Application Under Test (AUT) changes, these locators may no longer point to the intended elements, causing the test to break. Those changes can be categorized into two families: \textit{logical} and \textit{structural}~\cite{leotta2015using}.
A logical change modifies the AUT's behavior to introduce new features or alter existing ones, requiring updates to the test cases, but typically not to the locators themselves. A structural change, on the other hand, affects the layout or organization of the GUI without necessarily changing the underlying functionality, e.g., renaming an element ID or restructuring the DOM \cite{de2024investigating}. Such changes frequently invalidate existing locators, causing tests to break even when the application behavior remains correct. This phenomenon is known as a locator fragility or breakage~\cite{stocco2018visual}.

In this work, a modification to a test locator following the evolution of the AUT is called a locator change ($L_c$). A locator break ($L_b$) occurs when the previous locator can no longer identify the intended element in the updated application. Not every locator change is due to a break. A non-breaking locator change ($L_{nb}$) occurs when the original locator would still work, meaning the update was merely prophylactic. Every locator change thus belongs to one of these two categories, i.e., $L_c = L_b \cup L_{nb}$.
Figure~\ref{fig:example} illustrates such a scenario, where a locator used by Test~v1 to identify a button was renamed in App~v2, causing a breakage when Test~v1 is executed on App~v2. To determine which case applies, given a commit that modifies a locator, we first verify that the new locator passes on the updated application, and then reinsert the old locator into the same commit. If the test fails, we classify it as a $L_b$. If it passes, the change was not driven by a break ($L_{nb}$).
\begin{figure}[hbt!]
\begin{minipage}[t]{0.49\columnwidth}
\begin{lstlisting}[language=JavaScript, title=Test v1 / App v1]
// Test v1
cy.get('.btn-action')
    .should('be.visible');

// App v1
<button class="btn-action">
  Action
</button>
\end{lstlisting}
\end{minipage}
\hfill
\begin{minipage}[t]{0.49\columnwidth}
\begin{lstlisting}[language=JavaScript, title=Test v2 / App v2]
// Test v2
cy.get('.action-btn')
    .should('be.visible');

// App v2
<button class="action-btn">
  Action
</button>
\end{lstlisting}
\end{minipage}
\caption{Locator break example. Renaming the CSS class in App v2 invalidates the locator in Test v1.}
\label{fig:example}
\end{figure}

Locator fragility has been extensively documented in the literature. Christophe et al.~\cite{christophe2014prevalence} found that conventional locators caused up to 75\% of Selenium test files to change every nine commits. An analysis of 1,065 test breakages across 453 web application versions showed that fragile locators caused 73.6\% of the failures \cite{hammoudi2016record}. A systematic literature review \cite{nass2021many} further identified robust identification of web elements as the most prominent challenge in GUI test automation, and a recent study \cite{ricca2019three} pointed out that the creation of test scripts that are overly sensitive to small application changes is one of the three main problems in GUI testing.
Several techniques have been proposed to address this problem, including more resilient locator generation \cite{leotta2016robula}, similarity-based repair \cite{nass2023similarity, nass2023robust}, and NL-based testing~\cite{leotta2024empirical}. However, these were primarily developed and evaluated for Selenium, which sends commands to the browser through an \textit{external} driver. Cypress, on the other hand, runs directly \textit{inside} the browser alongside the application, and Playwright controls patched versions through \textit{internal} APIs~\cite{garcia2024exploring}. While all frameworks support CSS selectors and XPath, Cypress and Playwright also provide semantic locator APIs, such as \texttt{cy.contains}, and \texttt{page.getByText}.
%
To the best of our knowledge, no publicly available dataset of reproducible locator breaks exists for Cypress or Playwright. To fill this gap, we introduce \textbf{ReproBreak}, a dataset of reproducible locator breaks from open-source web applications using Cypress and Playwright for GUI testing. In summary, we make the following contributions:

\begin{enumerate}
    \item Analyzing 359 open-source repositories that use Playwright (189), Cypress (154), or both (16) for GUI testing to identify locator changes and breaks.
    \item Creating a structured dataset of locator changes and breaks.
    \item Providing scripts to automatically reproduce locator breaks, to facilitate further research and tool development.
\end{enumerate}

\section{Dataset Construction}\label{sec:cons}

\subsection{Data Source}

Our dataset is derived from the E2EGit dataset~\cite{meglio2025e2e}, which consists of 472 open-source web application projects, each containing a web GUI test suite implemented in one of the following frameworks: Cypress~\cite{cypress2026}, Playwright~\cite{playwright2026}, Puppeteer~\cite{puppeteer2026}, or Selenium~\cite{selenium2026}. From this dataset, we extracted GitHub repository links and filtered the projects to those that use Cypress or Playwright, resulting in 374 open-source repositories, 191 for Playwright and 183 for Cypress. Of these, 15 were no longer available on GitHub at the time of our analysis, leaving 359 repositories for our study. E2EGit was constructed by mining GitHub repositories and validated in prior work~\cite{di2025investigating}, making it a representative starting point.

\subsection{Data Collection}

\begin{figure*}[!tbp]
    \centering
    \includegraphics[width=0.80\textwidth]{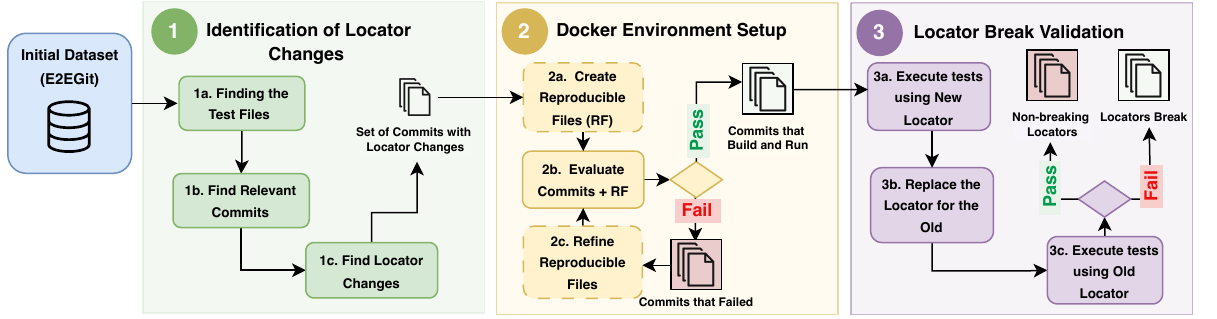}
    \caption{Data Collection Pipeline.
    Dashed boxes (step 2) indicate manual efforts.}
    \label{fig:diagram}
\end{figure*}

Our data collection pipeline (see Figure~\ref{fig:diagram}) consists of three steps: identification of locator changes, Docker-based environment setup, and locator break validation. The collection process is automated using a Python script, except for the creation of the reproduction files used in the Docker-based environment setup, which were crafted manually. The dataset cut-off date was April 21, 2026.

\subsubsection*{Identification of Locator Changes}

After cloning the 359 repositories, we identify all test files in the latest commit, following the same approach as E2EGit \cite{meglio2025e2e}. We filter by file extension to support only Java, JavaScript/TypeScript, and Python files. For each language, we apply rules that suggest test file formats, e.g., \texttt{.spec.ts} for TypeScript. When we find such a file, we read its content and apply a regex to detect framework-specific keywords, such as \texttt{cy.get} for Cypress or \texttt{page.locator} for Playwright, to confirm it contains web GUI tests.
Next, we parse the output of \texttt{git log} to identify which commits modified each test file, mapping each test file path to the commits that changed it. For each commit where a test file was modified, we store the commit hash, date, and previous commit hash, so we can later inspect both versions.
Finally, for each commit pair we run \texttt{git diff} to compute the differences, resulting in a list of pairs of code hunks that pinpoint the changes. Each hunk contains the old and new versions of a changed section. We apply regex patterns to detect locator usages in both versions. These patterns were designed to match both traditional selectors, such as \texttt{cy.get} in Cypress or \texttt{page.locator} in Playwright, and semantic APIs such as \texttt{page.getByRole}, \texttt{page.getByText}, and \texttt{cy.contains}, and extract their parameters. When we find a locator removal and the addition of another in consecutive lines within a hunk, we classify it as a locator change~($L_c$).
Note that locator calls whose selector argument spans multiple lines are not captured.

\subsubsection*{Docker Environment Setup}
To reproduce a locator break, we need to run both the application and the tests. Since each project has different dependencies, e.g., a local database like PostgreSQL or whole programming languages like PHP, we encapsulate everything in a Docker environment to ensure consistent and reproducible execution. Given the manual effort to set up these environments, we focused on the top-10 projects with the highest number of locator changes. From those, we were able to create reproduction files for 4/10 (see Table~\ref{tab:reproducible}). The other 6/10 could not be reproduced due to insufficient documentation, inaccessible dependencies, or unavailable environment configurations.
\begin{table}[t]
\centering
\begin{tabular}{l|cccc}
\hline
\textbf{Repository} & \textbf{\#$L_c$} & \textbf{\#$L_c^{valid}$} & \textbf{\#$L_b$} & \textbf{\#$L_{nb}$} \\ \hline
ghiscoding/angular-slickgrid & 344 & 270 & 258 & 12  \\
nasa/openmct                 & 271 & 88  & 48  & 40  \\
tryghost/koenig              & 264 & 155 & 93  & 62  \\
microsoft/playwright         & 399 & 67  & 50  & 17  \\ \hline
\textbf{Total}               & 1,278 & 580 & 449 & 131 \\ \hline
\end{tabular}
\caption{Number of locator changes (\#$L_c$), inspected (\#$L_c^{valid}$), locator breaks (\#$L_b$), and non-breaking (\#$L_{nb}$) per repository. \#$L_c^{valid}$ corresponds to commits that are validated.}
\label{tab:reproducible}
\end{table}
For each project, the reproduction files consist of a Dockerfile that sets up the application and its dependencies, and a script that starts the application and executes a given test file. These files are first created for the latest version of the application (step 2a).
The Docker image is built once and reused for all test files in that commit.
In step 2b, we evaluate the reproduction files against all identified commits. 
For failing builds or executions, we refine the reproduction files and repeat the process (step 2c) until no further changes can be made.

\subsubsection*{Locator Break Validation}
Having the locator changes and reproduction files, we validate each inspected $L_c$ (steps 3a--3c in Figure~\ref{fig:diagram}). In step 3a, we execute the test file using the new locator on the updated application. If the tests pass, we replace the new locator with the old one in step 3b and execute the tests again in step 3c. If the test now fails, we classify it as a locator break~($L_b$) because the only difference between the two executions is the locator itself. If it passes, we classify it as a non-breaking locator change~($L_{nb}$).

\section{Dataset Description}\label{sec:des}

\subsection{Dataset Overview}
\begin{figure}[H]
    \centering
    \includegraphics[width=0.42\textwidth]{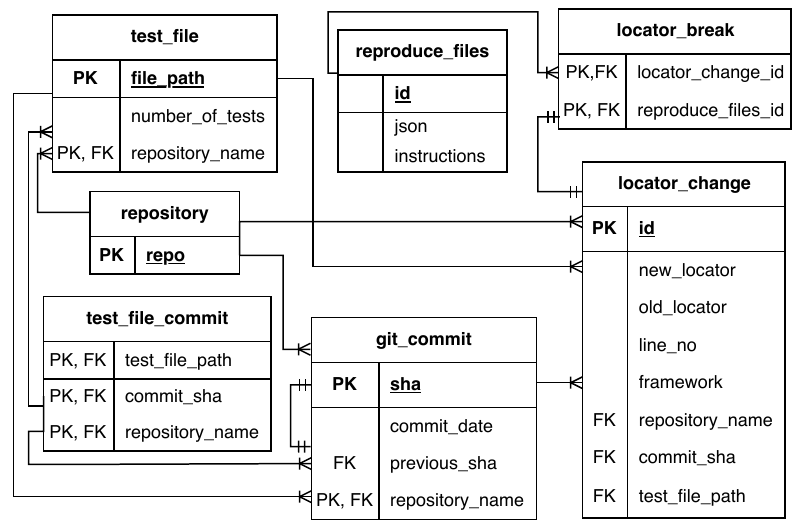}
    \caption{Entity-relationship model of the dataset.}
    \label{fig:db}
\end{figure}

The dataset is stored as a relational SQLite database. The schema (Figure~\ref{fig:db}) is centered around the \texttt{locator\_change} table, which stores the old and new locator, the line number, the framework, and references to the repository, commit, and test file where the change occurred. The \texttt{git\_commit} table captures the version history, storing each commit's hash, date, and a reference to its previous commit, so that any locator change can be traced back to both versions of the code. The \texttt{locator\_break} table connects a locator change to the reproduction files used to validate it, and the \texttt{reproduce\_files} table stores the Dockerfiles and shell scripts as JSON. The \texttt{repository} and \texttt{test\_file} tables store project-level and file-level metadata to support filtering and analysis, and \texttt{test\_file\_commit} records which test files were modified in each commit. 
Table~\ref{tab:summary} shows the number of locator changes identified from 211 out of the 359 repositories. Overall, we identified 9,572 locator changes.
%

\begin{table}[t]
\centering
\begin{tabular}{l|cc}
\hline
\textbf{Framework} & \textbf{$L_c$} & \textbf{Projects with $L_c$} \\ \hline
Playwright         & 4,186 & 111/189 (58.7\%) \\
Cypress            & 4,026 & 86/154 (55.8\%) \\
Both Frameworks    & 1,360 & 14/16 (87.5\%) \\ \hline
Overall            & 9,572 & 211/359 (58.7\%) \\
\end{tabular}
\caption{Summary of locator changes found per framework.}
\label{tab:summary}
\end{table}

\subsection{Reproduce a Locator Break}
The \texttt{reproduce.py} script retrieves files from the database, sets up the environment, and executes the tests. It takes a locator break \textit{ID} and one of three execution \textit{modes} as arguments (see Table \ref{tab:reproduction_modes}).
\begin{table}[t]
\centering
\small
\renewcommand{\arraystretch}{1.0}
\resizebox{\columnwidth}{!}{%
\begin{tabular}{c l l l}
\toprule
\textbf{Command} & \textbf{Mode \texttt{m}} & \textbf{Expected Result} \\
\midrule

\multirow{3}{*}{
\begin{tabular}{c}
\texttt{uv run reproduce.py} \\
\texttt{--locator\_id <id>} \\
\texttt{--mode <m>}
\end{tabular}
}

& fixed
& {\color{green}\checkmark} Test passed \\

\cmidrule(lr){2-3}

& reproduce\_break
& {\color{red}$\times$} Test failed \\

\cmidrule(lr){2-3}

& overwrite
& {\color{orange}?} Depends on changes \\

\bottomrule
\end{tabular}%
}
\caption{Reproduction script execution modes.}
\label{tab:reproduction_modes}
\end{table}
In \texttt{fixed} mode, the script executes the test file using the new locator on the updated application, and the test passes, confirming the valid locator change. In \texttt{reproduce\_break} mode, it reinserts the old locator into the same commit and reruns the tests, which fail, confirming the locator break.
The \texttt{overwrite} mode checks whether a custom repair resolves the break.
During the first execution, the Docker image for the specified commit is built automatically (see Figure \ref{fig:reproduce_script}); later the cached image is used. At the end, the script outputs the test file path, allowing to inspect the test and modify the locator.

\begin{figure}[t]
\begin{lstlisting}[
    language=ShellOutput,
    basicstyle=\ttfamily\footnotesize\color{black},
    breaklines=true,
    frame=single,
    numbers=none,
    xleftmargin=0pt,
    xrightmargin=0pt,
]
uv run reproduce.py --locator_id 1224 --mode reproduce_break
Getting info for locator break ID: 1224
  Repository: ghiscoding/angular-slickgrid
  Commit: b779ef14186481ebbd3ff32c1167729989feb3cf
Cloning repo: ghiscoding/angular-slickgrid
Extracting reproduce files from database...
  / Dockerfile
  / run_tests.sh
Setting up Docker image...
Docker image already exists. Skipping build.
Replacing locator to reproduce break...
Running GUI-tests tests...
Locator information
File path: /Users/../ReproBreak/data/...
Line number: 440
Locator change: cy.get('.slick-header-menu') -> 
    cy.get('.slick-header-menu .slick-menu-command-list')
Test failed!
\end{lstlisting}

\caption{Example of reproduce.py script execution.}
\label{fig:reproduce_script}
\end{figure}

\section{Application Scenarios}\label{sec:app}
ReproBreak enables more empirical studies and tool developments in web GUI test maintenance, as discussed below.

\paragraph{Locator Robustness Techniques.}
Approaches like Robula+~\cite{leotta2016robula} and Sidereal~\cite{leotta2021sidereal} derive more resilient locators, but have been built and evaluated exclusively for Selenium.
Now, ReproBreak enables their evaluation on Cypress and Playwright.

\paragraph{Fragility Score Assessment.}
The fragility score~\cite{di2024towards}, originally proposed and evaluated for Selenium, needs to be adapted for Cypress and Playwright, given the distinct locator strategies these frameworks use. Once adapted, ReproBreak can be used to evaluate it. Furthermore, analyzing which locator types present fewer breaks across the dataset can provide empirical grounding to refine the score and better capture what makes a locator break-prone.

\paragraph{Locator Repair Approaches.}
ReproBreak can be used to assess and compare locator repair approaches. E.g., techniques originally developed for Selenium, such as Similo~\cite{nass2023similarity}, Vista~\cite{stocco2018visual}, and Color~\cite{kirinuki2019color}, can now be evaluated on Cypress and Playwright projects. Additionally, AI agents for automated locator repair could be evaluated.

\paragraph{NL-Based Testing Robustness.}
ReproBreak can also be used to evaluate the robustness of NL-based testing techniques \cite{leotta2024empirical} regarding locator breaks. A potential direction could be to convert the test that uses the old locator into an NL-based representation and assess whether the abstraction level could have prevented the break. This was initially investigated on a smaller scale using Selenium \cite{kirinuki2022web}, and ReproBreak enables this evaluation at a larger scale.

\section{Conclusion}\label{sec:conc}
We presented ReproBreak, a dataset of reproducible locator breaks from open-source web applications using Cypress and Playwright. To the best of our knowledge, this is the first dataset of its kind for these frameworks. We analyzed 359 repositories, identified 9,572 locator changes, and reported 449 reproducible locator breaks across four projects. The scripts for automated reproduction make ReproBreak particularly valuable, allowing researchers to actively confirm breaks, test repair techniques, and benchmark their approaches in a real execution environment. ReproBreak can be extended to other frameworks by defining new locator regex patterns, and in the future, we plan to cover more projects and explore LLM-based agents to reduce the manual reproduction effort.


\section*{Data Availability Statement}
Our dataset and the corresponding code are openly available in our artifact:
\url{https://doi.org/10.5281/zenodo.21171968}~\cite{reprobreakDataset}.

\bibliographystyle{ACM-Reference-Format}
\bibliography{references}

\end{document}